\documentclass{ws-mpla}

\begin{document}

\markboth{Ishwaree P. Neupane} {Cosmic acceleration and M theory
cosmology}

%
\catchline{}{}{}{}{}
%

\title{Cosmic Acceleration and M Theory Cosmology}

\author{\footnotesize Ishwaree P. Neupane}

\address{Department of Physics, National Taiwan University, 106 Taipei,
Taiwan \\
Central Department of Physics, Tribhuvan
University, Kathmandu, Nepal \\
ishwaree@phys.ntu.edu.tw}



\maketitle

\pub{Received (24 January 2004)}{Published (Day Month Year)}

\begin{abstract}
This is a short overview of spatially flat (or open)
four-dimensional accelerating cosmologies for some simple
exponential potentials obtained by string or M theory
compactification on some non-trivial curved spaces, which may lead
to some striking results, e.g., the observed cosmic acceleration
and the scale of the dark energy from first principles.

\end{abstract}

\ccode{PACS Nos.: 98.80.Cq, 11.25.Mj, 11.25.Yb, 98.80.Jk}

\section{Introduction}

The universe appears to be accelerating, but the reason why is yet
to be explored. Cosmologists are trying to confirm the hypothesis
of a dark universe, providing two alternative candidates for the
dark energy -- a positive cosmological constant and a variable
$\Lambda$-term, like a slowly varying scalar potential; both exert
a gravitationally self-repulsive force. One would hold out less
hope of understanding the dark energy (or cosmic acceleration)
unless or until there is a unified theory with a compelling
particle-physics motivation that takes it closer to the bedrock of
space and time.

The usual idea in superstring or M theory is that spacetime is a
four dimensional nearly flat metric times a small six or seven
dimensional internal manifold. The current interest in cosmology
with the extra (spatial) dimensions is two fold : \\
(i) can inflation and/or observed cosmic acceleration naturally
arise from string or M theory compactification on some non-trivial
curved internal spaces of time-varying
volume~\cite{Townsend03a,Ohta03a,MNR1,IPN03b,IPN03c,IPN03d,Mohaupt},
explaining the scale of the dark energy~\cite{GKL03a,IPN03e}.\\
(ii) can one derive a scalar potential from string theory
compactification, with static and warped extra dimensions, that
has at least one stationary point with $V> 0$~\cite{KKLT}.

One of the obstacles for a warped de Sitter type compactification
is a no-go theorem~\cite{nogo1}. The strong energy condition holds
for all $D=10$ or $11$ supergravities. If one takes the extra
dimensions to be warped (and static), then for the compactified
theory (with or without form-fields in the extra dimensions), one
finds $R_{00}^{(4)} \geq 0$, which does not allow the universe to
accelerate. It seems less likely that a (false) de Sitter state
one may obtain in warped string theory background, with static
extra dimensions, is responsible for cosmic acceleration, because
such a vacuum is too unstable for a significant period of
inflation to occur. In certain cases, it may be possible to have
late time accelerated expansion from a tachyonic
potential~\cite{padma02a}.

The models like $V\sim \lambda \varphi^4$, $\sim m^2\varphi^2$, or
their hybrids, do not deal with the basic puzzles such as that of
the initial singularity, nor with the mystery behind dark energy
indicated by recent observation. So it might be worthwhile to
carefully consider the string or M theory compactification in time
dependent backgrounds. I shall argue that the dark energy of the
universe is possibly a gravitational scalar potential arising from
slowly varying size of extra dimensions, and so it is dynamical.

\section{Compactification with fluxes}

Let us recall that the four-dimensional part of a $(4+m)$
dimensional metric spacetime is the usual FLRW universe in the
standard coordinate (see e.g.,~\cite{IPN03c})
    \begin{equation}
      ds_4^2=g_{\mu\nu}(x) dx^\mu dx^\nu =
      -\,dt^2+a(t)^2\left(\frac{dr^2}{1-k r^2}+ r^2
      d\Omega_{2}^2\right),
    \end{equation}
    where $k=0,\,\pm 1$, $d\Omega^2$ the metric on a unit
    two-sphere, and $a(t)$ the
    time-dependent scale factor. This gives
    $R^{(4)}_{00}=-\,3{\ddot{a}(t)}/{a(t)}$ and
    inflating spacetime implies that $\ddot{a}/a>0$.
    The strong energy condition must be violated during
    inflation. To this end, we may allow the internal space to be
    time-dependent. As the simplest possibility,
    one can consider that the M theory spacetime is
    ${\cal M}_{4}\times \Sigma_{k_1,m}$:
\begin{equation}
 ds_{4+m}^2 = {\rm e}^{-m\phi(t)}
 g_{\mu\nu}(x)dx^\mu dx^\nu +
 r_c^2\,{\rm e}^{2\phi(t)}\,d\Sigma^2_{m, k_1}.
\end{equation}
with $\tilde{R}_{ab}(\Sigma_{k_1,m})=k_1(m-1)\tilde{g}_{ab}/r_c^2$
and $k_1=-1,0,+1$. For $k_1=-1$, $\Sigma={\cal H}^m/\Gamma$ is the
compact hyperbolic manifold (CHM). There is a mass gap and the
Kaluza-Klein excitations of the metric are heavy; $m_{KK}\gtrsim
r_c^{-1} \equiv M \simeq 10^{-3} M_P$ around $\phi\simeq 0$. The
field $\phi$ associated with the size and shape of extra
dimensions has an effective potential due to the curvature of the
internal space. With the 4-form field $F_{[4]}=2b
\mbox{vol}(\Sigma_{k_1,m})$, $b$ being the flux parameter, upon
the dimensional reduction we get~\cite{IPN03b,IPN03d,Emparan03a}
\begin{equation}
I= {M_P^2} \int d^4x \left(\frac{{\cal R}[g]}{2}- \frac{m(m+2)}{4}
\left(\partial\phi\right)^2 +\frac{m(m-1)k_1}{2 r_c^2} {\rm e}^{-
(m+2)\phi} - b^2 {\rm e}^{-3m\phi}\right),
\end{equation}
where $M_P=1/\sqrt{8\pi G}\equiv \kappa^{-1} $ is the reduced
Planck mass. The size and shape of the extra dimensions are
four-dimensional scalars. In the effective 4D theory, with a
canonically normalized scalar $\varphi$, one considers the
Lagrangian density
\begin{equation}
 {\cal L}= \sqrt{-g}\, \left(
      \frac{M_P^2}{2}\,{\cal R}[g]-(\partial\varphi)^2-2V(\varphi)
      \right).
 \end{equation}
The scalar potential $V(\varphi)$, for $k_1=-1$, takes the
following form
\begin{equation}
V(\varphi)=  {M_P^2 M^2}\,{\rm e}^{-\,2\,\lambda\,\kappa\varphi}
+\frac{1}{2}\, M_P^2 f^2 {\rm e}^{-\,({6}/{\lambda})
\kappa\varphi},
\end{equation}
where $f^2\equiv b^2 \left({4}/{m(m-1)}\right)^{3/c^2}$ and
$\lambda= \sqrt{\frac{m+2}{m}}$. $V(\varphi)$ could vary gradually
as the universe expands, which must be large during inflation,
where $\varphi=\varphi_0\simeq 0$ and $V\sim M_{P}^2 M^2$. It must
also be cancelled to extreme accuracy after
    inflation to allow the usual radiation and matter dominated eras,
    where $\varphi\gg {\varphi_0}$ and $V\sim 10^{-120} M_P^4$.

The equations of motion we need to solve are
    \begin{eqnarray}
&&    \ddot{\varphi}+3H\dot{\varphi}
    + \frac{dV}{d\varphi}=0 ,\label{wave1} \\
 &&  H^2+\frac{k}{a^2}=\frac{\kappa^2}{3}
 \left(\dot{\varphi}^2+2V(\varphi)\right).
    \end{eqnarray}
The curvature of the spatial part of the metric $k/a^2$ is related
to the total energy density of the universe, so this term cannot
be zero precisely. In any case, for $k=0$, it is convenient to
define a new logarithmic time variable $\tau$ by
$$d\tau={\rm e}^{-\lambda\,\varphi}\, dt\,, \quad
\alpha(\tau)=\ln(a(t)) ~\Rightarrow ~ a(t)=e^{\alpha(\tau)}.$$
With $b=0$, a class of expanding solutions (in units $M_P=1$)
is~\cite{IPN03e}
\begin{eqnarray}
{\sqrt{3}\alpha} &=& A_-\ln\cosh {M\gamma\tau}
+A_+\ln\sinh {M \gamma\tau}+ c_1 , \nonumber \\
 {\varphi} &=& A_-\ln\cosh {M\gamma \tau}-A_+
\ln\sinh {M\gamma \tau} + c_2,
\end{eqnarray}
up to a shift of $\tau$ around $\tau=0$, where $A_{\pm} =
(\sqrt{3}\pm \lambda)^{-1}\,, \quad
\gamma=\sqrt{(3-\lambda^2)/2}$. The solution found
in~\cite{Townsend03b} corresponds to $M=1$. One has
$H=\frac{da/dt}{a}={\rm e}^{-\lambda\varphi}\alpha^\prime >0$. The
critical value $\lambda=1$ separates qualitatively the different
cosmologies; the solutions with $\lambda>1$ are only transiently
accelerating (even if $b>0$), while, for $\lambda\leq 1$, the
accelerated expansion may be eternal, although $\lambda<1$ is not
obtainable from the compactification of classical supergravities;
the M-theory case is $\lambda=3/\sqrt{7}=1.13$

For $k=-1$, to lowest order, the zero-flux solution, with
$t>>M^{-1}$, is
\begin{equation}
a(t)=\frac{\lambda\,t}{\sqrt{\lambda^2-1}}
+\frac{\delta_1}{t^{\alpha}}=a_0+ a_1, \quad \varphi(t)=
\frac{1}{\lambda} \ln \left(M\lambda
t\right)+\left(\frac{a_1}{a_0}\right)
\frac{3(1-\alpha)\lambda}{4},
\end{equation}
where $\alpha^2=(2/\lambda)^2-3$. This corresponds to $-1 < w
<-1/3$ ($w\equiv p/\rho$). These are not inflationary solutions,
in the usual sense, specifically, if one is expecting $60$-efolds
or more, rather they may be the solutions responsible for the
current cosmic acceleration. It is healthy to keep in mind that,
for the late time cosmology (i.e., $t{\gg} M^{-1}$), there is no
need to solve the flatness (i.e., why $\Omega_\varphi \approx 1$)
and horizon problems. Rather, one should really look for such a
solution around $t \sim M^{-1}$, where $\varphi\simeq
\varphi_0\simeq 0$, with the scale factor $a(t)=H_0^{-1} \sinh
{H_0\,t}$, by satisfying $\Delta t\gtrsim 60 H_0^{-1}$.

For a non-zero field strength parameter $b$ (and/or a bulk
cosmological constant), which might serve as a source term, the
late time (i.e., $t>>M^{-1}$) cosmic acceleration can be eternal
for the coupling $\lambda <\sqrt{2}$, if $k=-1$. This observation
is not totally new, which was made before, but rather implicitly,
in the work of Halliwell~\cite{Halliwell87}, using a phase-plane
method, but the effects of flux were not considered there. For
$k=-1$, it is possible that there arise two periods of cosmic
acceleration. More discussions with two or more scalar fields will
appear in a separate publication.

\section{ Potential energy = cosmological constant?}

Is it possible to tune the M theory motivated potential to the
present value of the vacuum energy, so called cosmological
constant? This is indeed an interesting question. In the M theory
case $m=7$, and so $c=3/\sqrt{7}$, one has
\begin{equation}
V={M^2\,M_P^2} {\rm
e}^{-\frac{6}{\sqrt{7}}\,\frac{\varphi-\varphi_0}{M_P}}
+\frac{M_P^2\,f^2}{2}\,e^{-2\sqrt{7}\,\frac{\varphi-\varphi_0}{M_P}}
\end{equation}
The first exponent $6/\sqrt{7} \approx 2.27\equiv \lambda_*$ is
within the limit where astronomical data might be relevant,
$\lambda_* < \sqrt{6}$~\cite{Liddle98a}. If $M\sim 10^{-3} M_P\sim
10^{15}\,\mbox{GeV}$, in order to get $V(\varphi)\simeq 10^{-120}
M_P^4$, one requires $\varphi-\varphi_0\simeq 115$, in 4D Planck
units. Such a (large) shift in the value of $\varphi$ is not
impossible during the 13.7 billion years evolution of our
universe, but much of this shift should have been during inflation
and/or radiation dominated era. As the cosmology around $t=0$ is
not smoothly extrapolated to the present epoch ($t\sim
10^{17}\,\mbox{sec})$, the shift in $\varphi$ with $t$ need not be
uniform. The universe appeared to have gone through various phase
transitions, from the early pure vacuum dominated era
($\rho_{total}\equiv \rho_\varphi$) to the current era, where
$\rho_\varphi\simeq (2/3)\rho_{total}$.

For our purpose, it is rather important to know what is the factor
by which the size of the compact internal space may have grown up.
In terms of 4D metric, strictly speaking, to a 4D observer, the
growth factor (say $f_*$) is
\begin{equation} f_* =
\left(\frac{m(m-1)}{4}\right)^{1/(m+2)}
    \exp\left(\frac{2}{\sqrt{m(m+2)}}\,
    \frac{\varphi-\varphi_0}{M_P}\right).
    \end{equation}
For $m=7$ and $\varphi\simeq \varphi_0+115$, $f_* \sim 5\times
10^{12}$, one has $(f_* r_c)^{-1}\sim m_{KK}\sim 0.2~\mbox{TeV}$,
which is still interesting value. Let me elaborate a little on the
issue of size of the internal space (and its associated mass gap).
The authors in reference~\cite{GKL03a} argued that the $m_{KK}\sim
10^{-60}$ (in 4D Planck units), which is closer to the current
Hubble scale $H_0\sim 10^{-33}~\mbox{eV}$ and is
phenomenologically unacceptable. This estimation is rough and
presumably incorrect. The geometric bound on KK modes arising from
the hyperbolic compactification scales with $r_c^{-1}$ (or Ricci
curvature), other than the diameter, and not generically with the
volume of the manifold. For a tentative value of $M\sim
10^{15}\,\mbox{GeV}$, $m_{KK} \ll~\mbox{TeV}$ only if one allows
$V\approx V_0 <<10^{-120} M_P^4$. The lower bound for $m_{KK}$ may
be pushed up for the $b>0$ solution, by making $b\,r_c$ small. To
have a qualitative picture, consider the following solution, which
follows from $(4+m)$ dimensional field equations,
\begin{eqnarray}
&& ds_{4+m}^2 = {\rm e}^{-m\phi(T)}
\left(-\,S^{6}\,dT^2+S^2\,dx^2\right) + r_c^2\,{\rm
e}^{2\phi(T)}\,d{\cal H}^2_{m, k_1=-1},\\
&&\phi(T)= \frac{\ln(K\,L)}{m-1}, \quad S^2=
{K}^{{m}/{m-1}}\,{L}^{(m+2)/3(m-1)}, \\
&& K(T)= \frac{r_c}{(m-1)}\,\frac{\lambda_0\,\beta}
 {\sinh\left[\lambda_0\beta\, |T|\,\right]},\quad
L(T)= 2b\,\sqrt{\frac{m-1}{2m}}\,\frac{\cosh
    3\lambda_0\,T}{\lambda_0\,\beta},
\end{eqnarray}
up to a shift of $(4+m)$ dimensional coordinate time $T$ around
$T=0$, and $\beta\equiv \sqrt{\frac{3(m+2)}{m}}$. The constant
$\lambda_0$ has the dimension of inverse time. By suitably
choosing the flux parameter $b$ or the curvature radius $r_c$ (or
both), it is possible to suppress the growth in the size of the
internal space, so that all eigenmodes of Laplace-Beltrami
operator on a compact hyperbolic space have wavelengths less than
$f_*\times r_c$.

In summary, for a spatially flat FLRW universe, if dominated by an
exponential potential of a scalar field, $V=M^2M_{P}^2 {\rm e}^
{-2\lambda\kappa\varphi}$, the observed spell of cosmic
acceleration is only a transient phenomenon for $\lambda>1$, and
eternal for $\lambda\leq 1$. For all known classical
compactifications of 10 or 11d supergravities on some non-trivial
curved internal spaces (with or without fluxes) only $\lambda> 1$
arises in practice. So one is led to explore alternatives for the
observed cosmic acceleration. For $\lambda>1$, the cosmic
acceleration can be eternal if the curvature of the spatial
section of our universe is negative on large scales, although the
observable universe seems to be well described by flat Euclidean
geometry. It is plausible that the dark energy of the universe is
the gravitational scalar potential that arises naturally from
slowly varying size of extra dimensions. The spatial curvature
$k$, which might be negative, was significantly important during
the early-time inflationary periods.

\section*{Acknowledgments}

It is a pleasure to thank Pauchy W-Y. Hwang for organizing a
stimulating workshop, and Chiang-Mei Chen, Pei-Ming Ho and
Jiun-Huei Proty Wu for numerous helpful conversations. This work
was supported by the NSC and the CosPA project of the Ministry of
Education, Taiwan.

\end{document}